\documentclass{kapproc} 
\usepackage{procps}
\usepackage[dvips]{graphicx}
\upperandlowercase
\kluwerbib 
\begin{document}
\articletitle[Turbulence and galactic structure]
{Turbulence and galactic structure}
\author{Bruce G. Elmegreen}

\affil{IBM Research Division, T.J. Watson Research Center, 1101
Kitchawan Road, Yorktown Hts., NY 10598, USA}

\begin{abstract}
Interstellar turbulence is driven over a wide range of scales by
processes including spiral arm instabilities and supernovae, and
it affects the rate and morphology of star formation, energy
dissipation, and angular momentum transfer in galaxy disks.  Star
formation is initiated on large scales by gravitational
instabilities which control the overall rate through the long
dynamical time corresponding to the average ISM density. Stars
form at much higher densities than average, however, and at much
faster rates locally, so the slow average rate arises because the
fraction of the gas mass that forms stars at any one time is low,
$\sim10^{-4}$.  This low fraction is determined by turbulence
compression, and is apparently independent of specific cloud
formation processes which all operate at lower densities.
Turbulence compression also accounts for the formation of most
stars in clusters, along with the cluster mass spectrum, and it
gives a hierarchical distribution to the positions of these
clusters and to star-forming regions in general.  Turbulent
motions appear to be very fast in irregular galaxies at high
redshift, possibly having speeds equal to several tenths of the
rotation speed in view of the morphology of chain galaxies and
their face-on counterparts.  The origin of this turbulence is not
evident, but some of it could come from accretion onto the disk.
Such high turbulence could help drive an early epoch of gas inflow
through viscous torques in galaxies where spiral arms and bars are
weak. Such evolution may lead to bulge or bar formation, or to bar
re-formation if a previous bar dissolved. We show evidence that
the bar fraction is about constant with redshift out to $z=1$, and
model the formation and destruction rates of bars required to
achieve this constancy.  Bar dissolution has to be accompanied by rapid
bar reformation to get the constant bar fraction. This reformation is
consistent with numerical simulations by Block et al. (2002), but it
may not be possible according to models by Regan \& Teuben (2004).
The difference between these simulations is partly the result of a
difference in the two models for gas viscosity, which depends
on the approximations used to represent turbulence. In the Regan \&
Teuben model, a constant observed bar fraction implies that bars do not
dissolve significantly in a Hubble time.
\end{abstract}


\newpage

\noindent to appear in: Penetrating Bars through Masks of Cosmic
Dust: The Hubble Tuning Fork strikes a New Note, Eds., K. Freeman,
D. Block, I. Puerari, R. Groess, Dordrecht: Kluwer, in press
(presented at a conference in South Africa, June 7-12, 2004)

\section{Introduction}
Interstellar turbulence
plays a major role in the structure and dynamics of interstellar gas,
and through its influence on star formation, energy dissipation,
and viscosity, has a strong impact on galactic structure. Conversely,
galactic structure has an impact on turbulence through the energy that
comes from spiral instabilities.  A recent review of ISM turbulence is
in Elmegreen \& Scalo (2004) and Scalo \& Elmegreen (2004), and a review
of star formation in a turbulent medium is in Mac Low \& Klessen (2004).

Interstellar turbulence is difficult to model numerically because
a large number of resolution elements are required and there are
many terms in the equations. For example, Wada et al. (2002)
simulated the inner regions of a 2D disk with self gravity, star
formation and supernovae, but no magnetic fields. Shukurov et al.
(2004) did a 3D simulation with magnetic fields but no gravity. de
Avillez \& Breitschwerdt (2004) modeled the vertical structure in
a galaxy disk at high resolution.

Several points about ISM turbulence are important for this
discussion.

(1) The cool neutral medium is usually supersonic on scales larger
than several tenths of a parsec and therefore easily compressed in
the converging parts of turbulent flows.

(2) The magnetic field follows and resists this compression,
although shocks still occur parallel and perpendicular to the
field if the gas speed exceeds the Alfvén speed.  When the gas
moves slower than this, compression occurs mostly in the parallel
direction (e.g., Ostriker, Stone \& Gammie 2001).

(3) The energy density of motion in supersonic turbulence
dissipates rapidly, in about the crossing time of any flow region,
regardless of the presence of a magnetic field (Mac Low et al.
1998; Stone, Ostriker \& Gammie 1998; Mac Low 1999; Padoan \&
Nordlund 1999). This implies that bulk ISM velocities should
dissipate in $\sim20$ My, the flow time over the disk thickness,
and on much shorter times for smaller scales. Consequently, ISM
motions require constant stirring on the scale of the disk
thickness.

(4) Clouds, cloud cores, and other ISM structures, as well as
whole star-forming regions, should maintain their form and
activity for only one or two crossing times if their gas is
supersonically turbulent.  Clouds or cores where thermal motions
dominate might live much longer if they are not unstable to
gravitational collapse.

A crossing time corresponds to the shortest possible formation
time for stars, clusters, associations, and star complexes. Short
is in a relative sense though, not in the sense of an absolute
number of years, because the bigger structures take longer
absolute times to form.  Short formation times are measured
directly by the duration of star formation (Elmegreen \& Efremov
1996; Ballesteros-Paredes, Hartmann \& Vázquez-Semadeni 1999;
Elmegreen 2000; Hartmann et al. 2001). It follows that internal
energy sources are not required to generate the turbulence
observed in most clouds. It can usually be attributed to residual
energy left over from the compressive processes that formed the
clouds, in addition to gravitational binding energy for the
collapsing clouds.

Turbulence in {\it incompressible} flows can span a wide range of
velocities and spatial scales.  Fast flows on large scales contain
slower flows on smaller scales in a cascade that extends all the
way from the energy source down to the energy dissipation in
molecular collisions. The power spectrum of squared-velocity in
one dimension is close to a power law with an index of $-5/3$ for
incompressible flows.  This was predicted by Kolmogorov (1941)
using energy conservation for a downward cascade. The power
spectrum of any squared quantity in 1D is the energy spectrum,
$E(k)dk$. The energy spectrum is related to the power spectrum
$P(k)dk$ observed in higher dimensions $D$ by $E(k)dk=P(k)dk^D$.
This picture of a uniform cascade of turbulent energy in space and
velocity involves interactions between velocity wavenumbers that
are comparable in magnitude -- sometimes referred to as local in
wavenumber space. This is because large flows or eddies carry
along the smaller eddies, in a Galilean invariant sense, without
much interaction between the two. Possible complications from
non-locality in incompressible turbulence were discussed by Zhou,
Yeung \& Brasseur (1996).

(5) Supersonic ISM turbulence has non-local energy flow in shock
fronts when it carries energy from large scales directly to the
atomic mean free path without passing through an intermediate
cascade. Energy in supersonic turbulence also gets transferred
between solenoidal, compressional, and thermal modes. Boldyrev
(2002) predicted an energy spectrum $E(k)\sim k^{-1.74}$ by
assuming MHD turbulence is mostly solenoidal (incompressible) with
Kolmogorov scaling, and that the most dissipative structures are
shocks (see also Boldyrev, Nordlund \& Padoan 2002). Cho \&
Lazarian (2002) separated the shear (incompressible) and the fast
and slow compressible modes in a compressible MHD simulation for
the case where thermal pressure is much less than magnetic
pressure and found little coupling between the incompressible and
compressible parts.  For purely solenoidal driving, the shear
modes had $k^{-5/3}$ energy scaling for velocity and field, as did
the density in the weakly coupled slow mode; the weakly coupled
fast mode had $k^{-3/2}$ scaling for velocity, magnetic field, and
density.  Cho \& Lazarian (2003) got the same result for the high
pressure case.

(6) Supersonic turbulence can also carry energy from small to
large scales, as when an explosion produces a large moving shell.

There are many observations of power spectra for ISM column
densities and emission fluctuations (see review in Elmegreen \&
Scalo 2004). They are all approximately power laws with a slope of
around $-3\pm0.2$ in 2D maps (i.e., slightly steeper than the 2D
Kolmogorov slope of $-8/3$). These power laws often extend from
the smallest observable scale to the largest, including the whole
galaxy if data are available, as for the HI maps of the SMC
(Stanimirovic 1999) and LMC (Elmegreen, Kim, \& Staveley-Smith
2001).  Local HI emission (Dickey et al. 2001) shows the same
structure.

Another important difference between the Kolmogorov model of
incompressible turbulence and ISM turbulence is the spatial scale
for energy input.  In the Kolmogorov model, kinetic energy is
applied on some large scale and it causes motions on smaller and
smaller scales down to the dissipation length, where the advection
rate equals the dissipation rate. In contrast,

(7) ISM motions are stirred frequently and over a very wide range
of scales by various types of sources.

On large scales, turbulent energy comes from gravitational and
magnetic instabilities, gravitational scattering of nearby clouds,
cloud-disk impacts, and superbubbles. On intermediate scales it
comes from supernovae, stellar winds, and expanding HII regions.
On small scales it comes from low mass stellar winds and
gravitational wakes, Kelvin-Helmholtz and other fluid
instabilities, and possibly cosmic-ray streaming, although that is
probably more important in the ionized medium. Reviews of these
processes are in Norman \& Ferrara (1996), Mac Low \& Klessen
(2004), and Elmegreen \& Scalo (2004).

For the ISM, energy sources are so close together in time and
space that the gas reaction to one source of energy is usually
interrupted by another source before the first fully dissipates.
For example, a swing-amplified gravitational instability might
make a spiral arm and drive motions in the gas on a kpc scale, but
star formation inside this arm and self-gravity inside smaller
clouds will drive other motions on smaller scales before the
original spiral arm energy is dissipated.  This driving energy is
not necessarily partitioned into a power law in wavenumber space,
or at least not the same power law as the turbulent energy it
creates.  Thus the resulting power law for ISM turbulent energy
will in general be a combination of the distribution of scales for
the input energy and the distribution of scales for the non-linear
gas reaction to this input. Only on scales much smaller than the
smallest input, or for times that are significantly removed from
the last input event, can a state of pure gas turbulence be
realized. This might apply to scintillation observations, for
example (see Scalo \& Elmegreen 2004).

One observational implication of this multi-scale agitation is
that the ISM often resembles a network of shells or spiral arm
fragments with most of the cool neutral matter along the shells or
in the arms. Shells dominate the structure of the LMC (Kim et al.
1999) and other small galaxies (e.g., Ho II: Puche et al. 1992)
where shear is low, while shells (Brand \& Zealey 1975; Heiles
1979) and spiral arms tend to dominate the structure when shear is
high (high shear rate means in comparison to the shell growth rate
or the instability growth rate). The shells in the LMC are even
somewhat self-similar, spanning a range of scales over a factor of
at least $10$ (Elmegreen, Kim, \& Staveley-Smith 2001). Flocculent
spiral arms are self-similar too, combining into a power law power
spectrum (Elmegreen, Elmegreen, \& Leitner 2003). Thus shells and
spiral arms are intimately related to turbulence because both are
drivers of turbulence and both are structural reactions to
turbulence (Wada, Spaans, \& Kim 2000).

There is a similar multi-scale aspect to energy {\it dissipation}
in the ISM, making it different again from incompressible
turbulence:

(8) Energy dissipation in the neutral ISM covers a wide range of
scales, ranging from decompression regions downstream of spiral
arms or in disk-halo outflows, to ion-neutral slip viscosity in
all structures with field strength gradients, to continuous
dissipation in smooth magnetic shocks, to thin hydrodynamic
shocks. A similar range of scales is involved with dissipation of
turbulence in the ionized component (see review in Elmegreen \&
Scalo 2004).

The wide range of scales for both energy input and dissipation
underscores the difference between ISM turbulence and the standard
Kolmogorov picture of energy cascade from large scale input to
small scale dissipation.

In the next section, we outline several aspects of interstellar
turbulence as they are related to galactic structure. Given the
current level of uncertainty about the nature of ISM turbulence,
our view on many of these issues is rapidly evolving.

\section{Turbulence and Star Formation}

Stellar and galactic power sources in the ISM compress
interstellar gas and form clouds directly, often in shock fronts,
while shock-shock collisions and secondary shocks inside these
fronts make further structures down to very small scales (0.1 pc
or less).  The first generation shocks could be from a spiral
density wave or swing amplified instabilities, or it could be from
an explosion and subsequent shell. The secondary shocks are inside
the compressed clouds and in the hot cavities between them. If the
overall medium is significantly self-gravitating or if a cloud is
significantly self-gravitating, then some of the secondary shocks
inside these regions can produce gas that is also significantly
self-gravitating, and this gas can collapse into smaller clouds,
cloud cores, or stars before the ambient turbulence shears and
distorts the region out of existence. A review of these
star-formation processes is in MacLow \& Klessen (2004), while
simulations are in Gammie et al. (2003), Y. Li et al. (2003), P.S.
Li et al. (2004) and elsewhere. A different type of simulation is
in papers by Bate and collaborators (e.g., Bate, Bonnell \& Bromm
2003), who find collapse in turbulent gas down to very small
scales and masses where the optical depth becomes large. Then star
formation proceeds upwards in mass from there, as a result of
accretion.  In either case, cloud turbulence defines the primary
cloud structure and self-gravity inside this structure leads to
star formation.

An important result of turbulence simulations in isothermal gas,
as might apply to molecular cloud cores, is that the probability
distribution function (pdf) for density is approximately a
log-normal (e.g., Li, et al. 2004), with a power law tail in
gravitationally collapsing regions at high density (Klessen 2000).
The log-normal indicates that the gas density changes randomly and
multiplicatively by successive compressions and rarefactions
(Vázquez-Semadeni 1994).  When the gas becomes sufficiently
self-gravitating, this random, two-directional process stops and
the gas density increases monotonically until a star forms.
Observations suggest that the point of no return occurs at a
density of about $10^5$ molecules cm$^{-3}$. This may be just a
coincidence for the local regions that are usually observed, or it
may be the result of specific processes that change at this
density, promoting collapse. For example, at around $10^5$
cm$^{-3}$, big charged grains start to decouple from the magnetic
field, molecules freeze onto grains, and the turbulent speed drops
to about the sound speed, making further compressions difficult
(see review in Elmegreen 2000).

The density pdf may be integrated over the high density tail to
find the volume fraction of gas at high density, $f_V$.  In the
log-normal of Wada \& Norman (2001), which may
be representative of galaxy disks, the volume fraction at a
density above $10^5$ times the average density is equal to
$f_V=10^{-9}$.  Similarly, the mass fraction above $10^5$ times
the average density is $f_M\sim10^{-4}$. If turbulence establishes
the density sub-structure in clouds, and if gas collapses only at
densities greater than $10^5$ times the average, then this
$10^{-4}$ should be the fraction of the ISM mass that goes into
stars in each turbulent crossing time on the small scale, where
the collapse occurs. The star formation rate per unit volume on a
galactic scale is then
\begin{equation}
SFR/V=f_V\epsilon\rho_5\left(G\rho_5\right)^{1/2},\end{equation}
where $\epsilon$ is the fraction of the gas mass inside a dense
clump that goes into a star or stars in a dynamical time, $f_V$ is
again the volume fraction of the whole ISM above this density, and
$\rho_5$ is the high density where collapse becomes inevitable,
taken here to be $10^5\mu$ cm$^{-3}$ for mean molecular weight
$\mu\sim4\times10^{-24}$ g. The basic rate of star formation used
in this equation is the dynamical rate from gravity,
$\left(G\rho\right)^{1/2}$. Star formation usually proceeds at
about this rate with fairly high efficiency in a cloud core,
$\epsilon\sim0.3-0.5$ (e.g., Matzner \& McKee 2000). Setting the
density at $10^5$ cm$^{-3}$ and using $f_V=10^{-9}$ gives a galaxy
wide average star formation rate of $10^{-5}$ M$_\odot$ pc$^{-3}$
My$^{-1}$.

This is the same rate that comes from the Kennicutt (1998)
observation, which gives a star formation rate per unit area of
$SFR/A\sim0.033\Sigma\Omega$ for mass column density $\Sigma$ and
galaxy rotation rate $\Omega$. This observation applies to all
galaxies in Kennicutt's survey, to the inner parts of these
galaxies and to starburst galaxies. To convert this to a rate per
unit volume, we use the fact that the local density is always
about the tidal density, $\rho_{tid}=3\Omega^2/\left(2\pi
G\right)$, assume a flat rotation curve, and assume an exponential
disk. Then integrating over the disk from the center to an outer
edge at four exponential scale lengths gives an average star
formation rate per unit volume of
\begin{equation}SFR/V\sim0.012\rho\left(G\rho\right)^{1/2}
\end{equation}
The result for average density $\rho\sim1\mu$ g cm$^{-3}$ is the
same as the estimate above, $10^{-5}$ M$_\odot$ pc$^{-3}$
My$^{-1}$.

This exercise suggests that the star formation rate in essentially
all galaxies, averaged over large enough areas, is determined by
the available gas (the $\rho$ term) turning into stars on the
local dynamical rate, $\left(G\rho\right)^{1/2}$, with an
efficiency of $\sim0.01$.  This low average efficiency is
understood for a turbulent medium to be the result of turbulent
fragmentation: a small but predictable fraction of the gas is at a
density high enough to form stars, which is about $10^5$ times the
average $\rho$. The rest of the gas has a density lower than this,
making it more easily distorted by random turbulent flows.

\begin{figure}[ht]
\vskip.2in
\centerline{\includegraphics[width=5in]{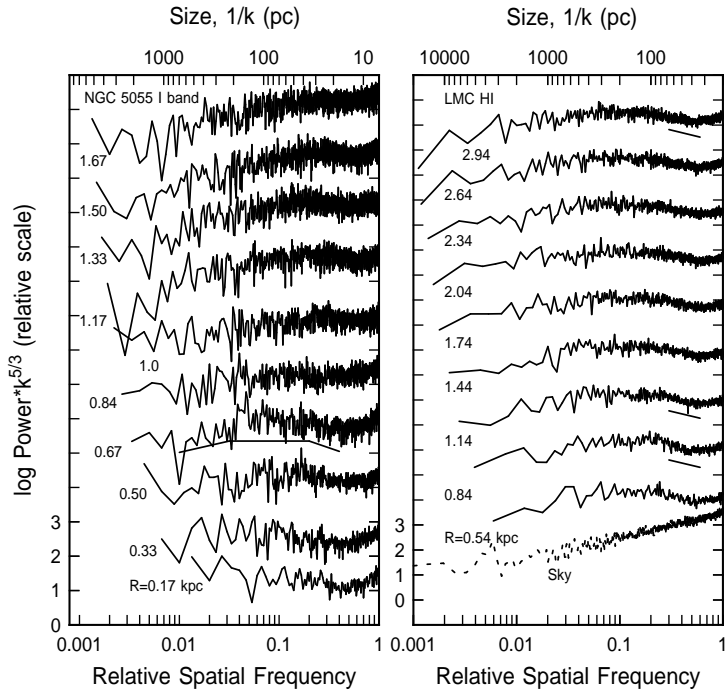}}
\caption{Power spectra of the azimuthal profiles of HI emission
from the LMC (right) and optical I-band emission from NGC 5055
(left). The power spectra are multiplied by $k^{5/3}$ to flatten
them for easy inspection if they are near the Kolmogorov spectrum
of $k^{-5/3}$.  The LMC gas and NGC 5055 star formation have
remarkably similar power spectra, and power spectra that are also
similar to that of incompressible turbulence.  This result
suggests that star formation distributions are regulated by
turbulent processes. (Figure from Elmegreen 2004.)}\label{fig:ps}
\end{figure}
\noindent

Turbulence not only partitions the gas into clumps, but it also
locates these clumps in a certain fashion, giving the overall ISM
a hierarchical structure with large clumps containing small clumps
in many levels of the hierarchy.  If stars form in the densest
regions of this gas, then they should have the same hierarchical
structure. This is widely observed to be the case. Zhang, Fall, \&
Whitmore (2001) showed that clusters in the Antenna galaxy are
correlated in a power-law fashion (i.e., hierarchically clumped
together) for scales less than $\sim1$ kpc. Elmegreen \& Elmegreen
(2001) found fractal structure in the star fields of many galaxies
and showed that the fractal dimension is about the same as for the
gas. Earlier work by Feitzinger \& Braunsfurth (1984), Feitzinger
\& Galinski (1987), Elmegreen \& Efremov (1996), and Efremov \&
Elmegreen (1998) also found fractal patterns for young stars.

Entire galaxies also have correlated optical structure from a
combination of star formation and dust extinction. Power spectra
of azimuthal profiles of the optical light from galaxies show
power-law forms with slopes comparable to $-5/3$ (Elmegreen,
Elmegreen \& Leitner 2003), which is the slope for the velocity
power spectrum in Kolmogorov turbulence.  The power spectrum of
the optical light in NGC 5055 is essentially the same as the power
spectrum of HI in the LMC, as shown in Figure \ref{fig:ps}. Models
including randomly positioned foreground stars, hierarchically
distributed clusters in the galaxy, plus hierarchically
distributed field-star light in the galaxy reproduce the
observations well (Elmegreen et al. 2003).

Different nomenclature is often applied to various parts of this
hierarchy, ranging from star complexes such as Gould's Belt on the
largest scales (Efremov 1995) to OB associations and subgroups on
smaller scales. These regions are all likely to be part of the
same physical processes involving gravitational instabilities,
turbulent fragmentation, and direct compression from stellar
pressures.

Recall the discussion in the introduction where the driving
sources for ISM turbulence were noted to be widespread and
overlapping in space and time. This means that shell formation and
cloud compression by high pressure events are sources for
turbulence as well as sources for triggering star formation. Star
formation operates only in the densest parts of the ISM, and all
of these stirring processes occur at much lower densities.  Thus
the organization to star formation on large scales may not be
important for the overall star formation rate. Nevertheless this
stirring affects the spatial distribution of the clusters that
eventually form, often placing them along the rims of shells or in
comet-heads.

Elmegreen (2002) summarized this dichotomy by saying that
turbulence fragments the gas into dense pieces, but these pieces
are constantly buffeted and compressed by the same energy sources
that drive this turbulence, triggering star formation in a high
fraction of cases.  As long at this turbulence makes a regular
density pdf, the overall star formation rate is determined in
large part by the fraction of the gas at high density. The rate
that comes from this fraction is about the same as the large-scale
rate that comes from gravitational instabilities at the average
density of the ISM. The details of the stirring
and local triggering processes are relatively unimportant (see
also Wada \& Norman 2001).

Another way to visualize this is to consider the bottleneck to
star formation in galaxies. The rate-limiting processes are all on
the large scale where the density is low and the dynamical time
long. On this scale, the onset of star formation is primarily by
gravitational processes, such as swing-amplified instabilities.
However, stars actually form on a much smaller scale. The link
between these two scales is provided by turbulence.   The small
dense clumps made by this turbulence each turn into stars very
quickly, and have virtually no consequence, individually, to the
processes on the large scale because they represent only a very
small fraction of the total gas mass ($10^{-4}$). However, these
small clumps are continuously regenerated by the turbulence on
larger scales, forming more stars, and this regeneration continues
for the dynamical time on the large scale. In each dynamical time
on the large scale, 100 cycles of core regeneration occur, turning
$100\times10^{-4}=1$\% of the gas into stars.  This gives the
Kennicutt (1998) star formation law for all galactic regions,
including starbursts.

\section{Turbulence and Disk Accretion}

Turbulence produces viscosity, which leads to disk accretion. If
the viscous time is proportional to the star formation time, then
a gas disk can evolve toward an exponential profile (Lin \&
Pringle 1987; Yoshii \& Sommer-Larsen 1989; Saio \& Yoshii 1990;
Gnedin, Goodman \& Frei 1995; Ferguson \& Clarke 2001).  This
process does not appear to be self-regulating, however.  If the
viscous time is much less than the star formation time, then there
is the type of feedback that is needed for regulation: the disk
accretes, the density increases, the star formation rate per unit
area increases, and the two rates come into balance.  However, if
the viscous time is much greater than the star formation time,
then there is no feedback: star formation simply removes the gas
by turning it into stars, and the accretion stops.  Modern
cosmology simulations produce exponential disks from the start
(Robertson et al. 2004), without needing an evolutionary process
involving accretion and star formation, so viscous production of
exponential disks is not necessary.

Disk accretion brings gas to the centers of galaxies, leading to
bar destruction if the accreted gas mass is large enough (Hasan \&
Norman 1990; Pfenniger \& Norman 1990; Bournaud \& Combes 2002;
Debattista et al. 2004).  Negative torques from viscous accretion
outside the bar region are usually overcome by positive bar and
spiral arm torques there, which drive a net outflow. Accretion
inside the bar region is driven mostly by bar torques, with
positive or negative pressure gradients that contribute to these
torques. The net accretion rate depends on the energy loss in
addition to the torques.  This energy loss is uncertain but likely
to be rapid. Energy from star formation can restore some of the
internal gas energy downstream, and the pressure from this energy,
pushing against the front side of the spiral or bar, can restore
some of the lost angular momentum.

The equation of motion for a fluid has a contribution to the time
derivative from viscosity that can be written $\nu\nabla^2 v$,
from which the accretion time may be estimated to be $D^2/\nu$ for
inverse gradient distance $D$. The viscous coefficient $\nu$ is
from turbulence rather than molecular collisions, and its value is
unknown. If it can be represented by the product of a length and a
speed, $\nu=\lambda c$, then $\lambda$ might be the outer scale
for the correlated motions and $c$ the rms turbulent speed on that
scale. Both of these quantities are highly uncertain, particularly
because the ISM may have a 3D type of turbulence on scales smaller
than the disk thickness and a 2D turbulence on larger scales
(Elmegreen, Kim, \& Staveley-Smith 2001). A good guess for
$\lambda$ might be the disk thickness itself, in which case
$\lambda\sim200$ pc. Then the velocity dispersion is the rms speed
of most of the gas mass, which is $\sim5$ km s$^{-1}$. These
parameters give an accretion time over distance $D$:
\begin{equation}t_{acc}\sim{{10\;{\rm Gy}\left(D/{\rm kpc}\right)^2}\over
{\left(\lambda/200\;{\rm pc}\right) \left(c/5\;{\rm km
\;s}^{-1}\right)}}.\end{equation} This is a very long time, even
longer if we consider that the real length for the disk gradient
is the exponential scale length, which is typically several kpc.
Numerical simulations of galaxy disks with gas represented by
discrete particles that have longer mean free paths than the disk
thickness (averaged over an orbit) can have shorter viscous
accretion times than this, possibly leading to spurious effects.

Given the likely small value for the viscous coefficient, disk
accretion is dominated by gravitational torques produced in spiral
arms and bars. Spirals and bars transfer angular momentum from the
inner disk to the outer disk (Lynden Bell \& Kalnajs 1972).
Bars produce accretion in the inner part, often to a
nuclear ring (Regan \&
Teuben 2004), and outflow in the outer part, often to an outer
resonance ring (Schwarz 1981). Bars are much stronger perturbers
than spirals because typically only barred galaxies have outer
resonance rings (Buta \& Combes 1996). The lack of outer resonance
rings in non-barred galaxies seems to imply that these galaxies
never had a significant bar in their past.  This may imply there
are relatively few galaxies that have dissolved bars (see Section
\ref{sect:bar}).

\section{Turbulence, Viscosity, and Bar Dissolution}
\label{sect:bar}

The evolution of galaxies over a Hubble time has been studied
extensively by simulations but has only recently been observed
directly through deep images with the Hubble Space Telescope.
Young galaxies often appear physically small and at high restframe
surface brightness (Bouwens \& Silk 2002), although cosmological
dimming allows us to see only the highest surface brightness
members of a sample. Galaxies with normal sizes are also present
at high z (Simard et al. 1999; Ravindranath et al. 2004). Of
interest is the process of galaxy growth and the redistribution of
mass inside galaxies during growth.  Turbulent viscosity and
dissipation play important roles in this redistribution.

Chain galaxies (Cowie, Hu, \& Songalia 1995) are interesting
because they appear to be unique to high z.  They are linear
structures with several large bright clumps and no exponential
disk or bulge. If they are edge-on disk galaxies (Dalcanton \&
Schectman 1996; Reshetnikov, Dettmar, \& Combes 2003; Elmegreen,
Elmegreen, \& Sheets 2004a; Elmegreen, Elmegreen, \& Hirst 2004b),
then their local analogues do not have clumps that extend nearly
as far in the vertical direction (e.g., Hoopes, Walterbos, \& Rand
1999). The large clump size implies high-speed turbulent motions
if the clumps are self-gravitating. The lack of spirals in the
face-on counterparts implies that turbulence dominates shear
during star formation (Elmegreen, et al. 2004b).

\begin{figure}[ht]
\vskip.2in
\centerline{\includegraphics[width=4.5in]{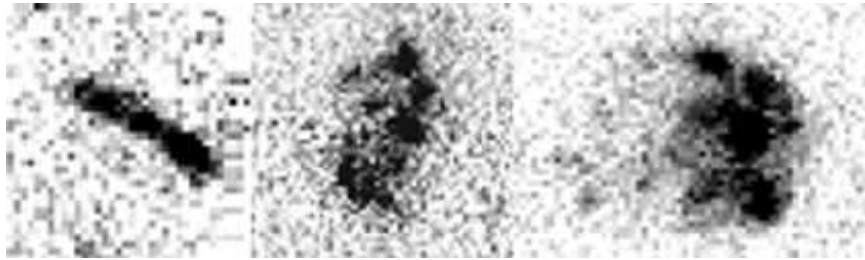}}
\caption{Three high-redshift galaxies from the background field of
the Tadpole galaxy are juxtaposed to suggest that chain galaxies
(like that on the left) are edge-on versions of clump clusters
(like that on the right). This projection is confirmed by the
distribution of width-to-length ratios, which is flat as in normal
disk galaxies. Neither chains nor clump clusters have exponential
disks or prominent red bulges near their centers. Irregular high-z
galaxies like this have giant blue clumps that suggest star
formation occurred in gas rich disks with large turbulent speeds,
comparable to several tenths of the rotation speed. Turbulence in
young galaxies could be the result of high galactic accretion
rates. (From Elmegreen et al. 2004b.)}\label{fig:cc}
\end{figure}
\noindent

Figure \ref{fig:cc} shows three galaxies in the deep field of the
Tadpole galaxy (Tran et al. 2003) where we found 69 chain galaxies
with three or more giant clumps (as shown on the left in the
figure), 58 other linear structures with one or two clumps, and 87
tight clusters of clumps that looked like face-on versions of the
chain galaxies (as shown in the middle and right frames of the
figure). None of these objects have exponential disks or bright
red clumps in their centers that could be bulges.  The colors and
magnitudes of the clumps and of the whole galaxies in these
samples are all about the same, and the distribution of the
width-to-length ratio is flat down to a lower limit of $\sim0.2$.
Such a flat distribution is appropriate for circular disks and
similar to that for local spiral galaxies in the RC3 (de
Vaucouleurs et al. 1991; Elmegreen et al. 2004b). The implication
of these results is that clump-clusters are probably face-on
versions of chain galaxies.

\begin{figure}[ht]
\vskip.2in
\centerline{\includegraphics[width=5in]{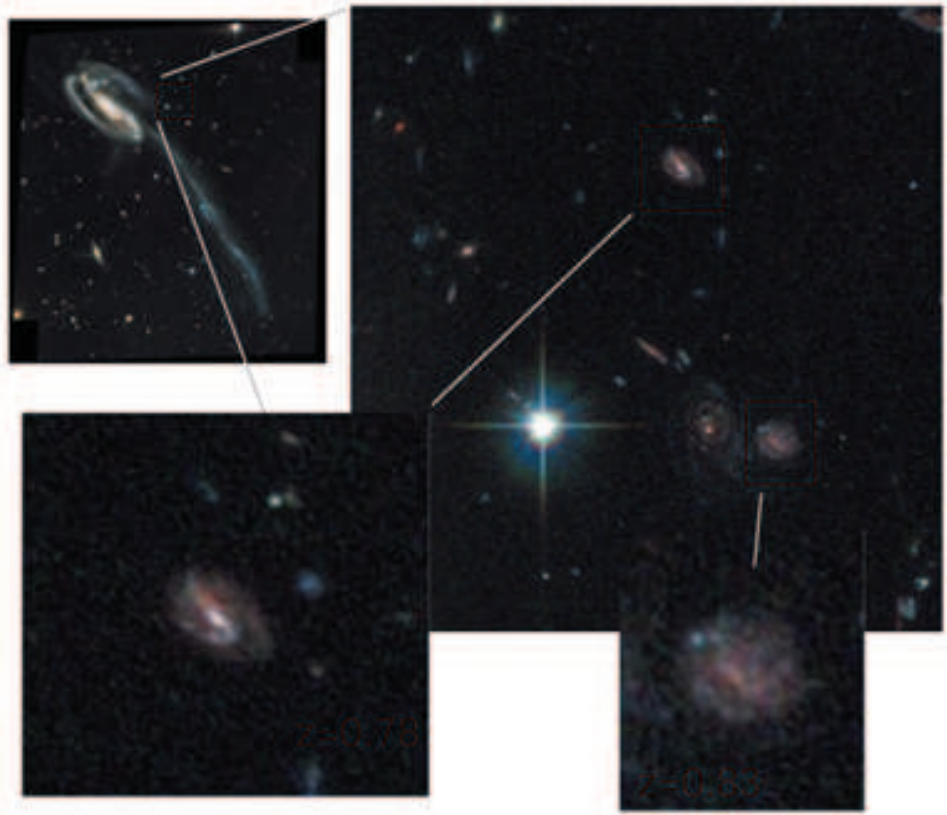}}
\caption{The Hubble Space Telescope Advanced Camera for Surveys
field of the Tadpole galaxy (Tran et al. 2003) is shown in two
successive stages of blow-ups revealing normal-looking barred
galaxies.  This field contains 22 clear barred galaxies like this
and 21 additional barred galaxies that are so small their bars
show up mostly as inner twists in isophotal contours. The
abundance of bars at high z is important for studies of bar
dissolution following gas accretion. (for higher resolution, see
the jpg version in astro-ph.)}\label{fig:twobars}
\end{figure}
\noindent

The giant clumps in these galaxies are blue and likely to be
star-forming regions.  Their diameters are $\sim500$ pc and they
are spaced from each other by several kpc, which is several tenths
of the disk diameter.  The fact that there are just a few giant
clumps per galaxy, along with their dominance of the disk light
and the lack of obvious spiral arms, suggests they formed by
gravitational instabilities in a medium that is mostly gas and has
a relatively high turbulent speed (Noguchi 1999; Immeli et al.
2003, 2004). If we set the Jeans length $1/k_J=c^2/\left(\pi G
\Sigma\right)$ equal to $1/4$ the galaxy radius, $R$, for gas
surface density $\Sigma$, then we need a turbulent speed
$c\sim2V\left(\Sigma/\Sigma_T\right)\sim0.3V$ to $0.5V$ for orbit
speed $V$, where $\Sigma_T$ is the total effective surface
density, including dark matter, that contributes to the rotation.
At this turbulent speed, the instability time is comparable to the
orbit time and collapsing regions should be spun up by Coriolis
forces unless there is a magnetic field in the disk. The
brightness of the clumps compared to the underlying disk suggests
they are among the first generations of star formation. Presumably
some will eventually merge to form an exponential disk or a bulge
(Noguchi 1999).

\begin{figure}[ht]
\vskip.2in
\centerline{\includegraphics[width=5in]{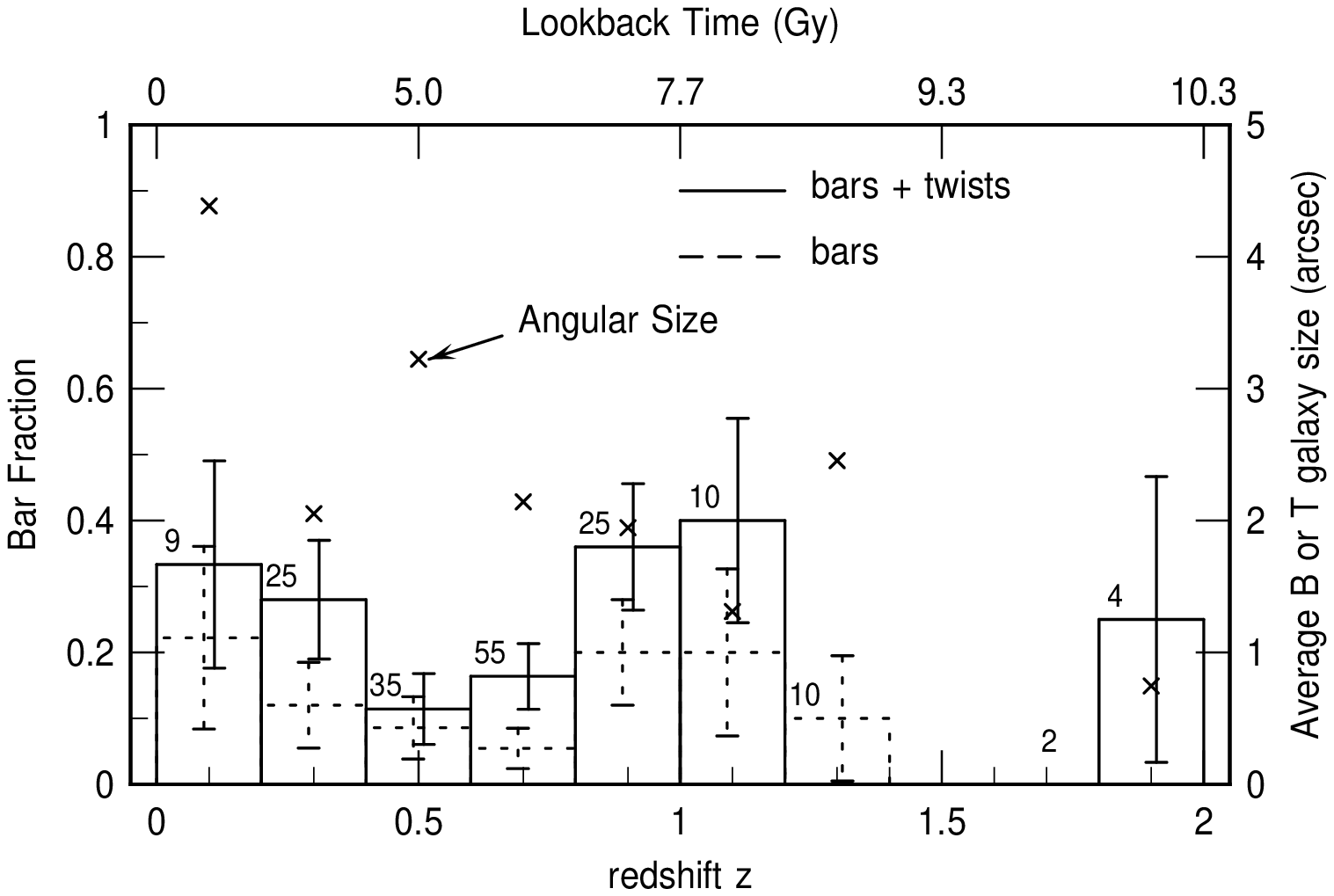}}
\caption{The bar fraction is shown as a function of redshift z for
galaxies in the Tadpole field. The solid line histograms are for
all bars in the sample, including those which show up only on
contour plots. The dashed lines are for the clearest cases of
bars. The bar fraction is about constant, suggesting either that
bar dissolution is unimportant or bars regenerate relatively
quickly once they dissolve.  Bar regeneration implies that
accreted gas in the outer disk can make its way to the center
where it can drive a new bar instability, according to Block et
al. (2002). Such accretion without a bar depends on turbulent
viscosity and on gravitational torques from spiral
arms. (From Elmegreen et al. 2004c.)}\label{fig:bf}
\end{figure}
\noindent

The origin of the turbulence in these galaxies is not clear.
Shells or other reactions to star formation are not evident, so
most of the turbulent energy may come from the galaxy formation
process itself.

Barred galaxies at high z are also prominent in this field. We
found 22 clearly barred galaxies out to $z=1$, complete with grand
design spiral arms, bulges, and exponential disks. There were also
another 21 galaxies that looked barred on contour plots, which
showed an inner isophotal twist.  The bar fraction was determined
as a function of the ratio of axes and compared with the local bar
fraction.  Bars were less prominent at high inclinations, more so
than local bars at equally high inclinations; the difference is
probably the result of poor resolution for the distant bars. We
estimated that perhaps twice as many bars were lost to inclination
effects at high z than locally.  The bar fraction was also
determined as a function of $z$ using photometric redshifts from
Benitez et al. (2004).  This fraction is about constant out to
$z\sim1$ and equal to 0.2 to 0.3. Corrected for inclination, the
bar fraction is about the same as the local fraction, which is
$\sim0.4$ in B band depending on Hubble type (Elmegreen et al. 2004c).
Figure \ref{fig:twobars} shows two bars in successive blow-ups of
the Tadpole field.  Figure \ref{fig:bf} shows the bar fraction as
a function of redshift.

We suggested that a constant bar fraction with $z$ over the last
$\sim 8$ Gy (out to $z=1$) offers no evidence for bar dissolution
over a Hubble time (Elmegreen et al. 2004c).  If bars dissolved,
then they had to reform, as suggested by Block et al. (2002).
However, if non-merger interactions preferentially formed bars,
rather than destroyed them (Noguchi 1987; Gerin, Combes \&
Athanassoula 1990; Berentzen et al. 2004), and if such
interactions were more frequent in the past because of the higher
galaxy density, as is likely, then the bar formation rate was
higher in the past. To maintain a near-constant bar fraction, this
means either that the dissolution rate had to be much higher in
the past, or that most bars formed early in the Universe and did
not dissolve.  Regan \& Teuben (2004) suggest, for example, that
gas accretion in a bar stops at an ILR ring and does not get to
the center, in which case the bar would not dissolve.

\begin{figure}[ht]
\vskip.2in
\centerline{\includegraphics[width=4in]{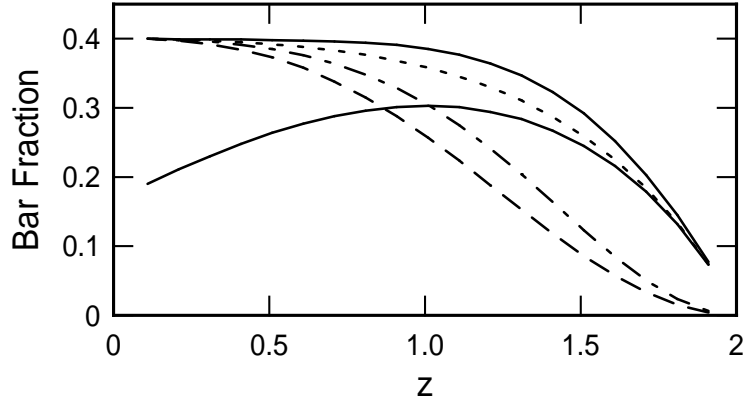}}
\caption{A simple model of bar formation and destruction by
various processes is shown to illustrate how prompt bar
reformation following internal dissolution can maintain a constant
bar fraction out to $z\sim1$. The dotted line forms bars by galaxy
interactions from $z=2$ to today and has no bar dissolution. The
dashed line has bar formation by internal processes with a rate
that increases with time at first and then decreases exponentially
with an e-folding time of 0.1 Hubble time. The dot-dashed line has
the same internal process with a bar dissolution rate proportional
to the formation rate.  The solid line with a constant bar
fraction out to $z=1$ includes all three processes. All of these
examples were tuned to give a bar fraction of 0.4 today.  The
solid line with a bar fraction that increases out to $z=1$ has bar
formation by interactions and bar dissolution by internal
processes with a time scale of 10 Gy. Faster dissolution makes the
rise to $z=1$ larger.  This latter case illustrates that even a
small amount of bar dissolution cannot operate without continuous
re-formation if the bar fraction is to be constant out to $z=1$.}
\label{fig:deepbar2}
\end{figure}
\noindent

A simple model illustrates how bar dissolution is possible if the
dissolution rate is proportional to the internal formation rate
and an additional formation rate from collisions is proportional
to the square of the co-moving density out to $z=2$.  Suppose bars
form by collisions at the rate
\begin{equation} {\cal F}_{col}={\cal
I}_0\left(1+z\right)^6/\left(1+2\right)^6
\end{equation}
for constant ${\cal I}_0$.  Here we normalize to the rate at
$z=2$. If this were the only bar formation process and there were
no bar destruction, then ${\cal I}_0=0.51$ Gy$^{-1}$ gives a bar
fraction of $f=0.4$ today. Suppose bars also form by internal
processes at a rate given by
\begin{equation}
{\cal F}_{int}\left(z\right)={{{\cal F}_0\left(t-t_2\right)}\over
{0.1\tau}}e^{\left(t-t_2\right)/\left(0.1\tau\right)}.
\end{equation}
This formation rate increases at first and then decreases with an
exponential time scale of 0.l times the age of the Universe,
$\tau$; $t_2$ is the time at $z=2$. If this were the only bar
formation process, then ${\cal F}_0=0.385$ Gy$^{-1}$ gives a bar
fraction of $f=0.4$ today. Finally, suppose the bar destruction
rate is proportional to the internal formation rate:
\begin{equation}{\cal F}_{des}={\cal D}_0{\cal
F}_{int}.\end{equation} If bars only formed at the internal
process with ${\cal F}_{int}=0.385$ Gy$^{-1}$, and if ${\cal
D}_0=1$, then the final bar fraction would be 0.32 with 9\% of all
formed bars having been destroyed.

To solve for the bar fraction $f$ we use the equation
\begin{equation}
{{df}\over{dt}}={\cal F}_{col}(1-f)+{\cal F}_{int}(1-f)-{\cal
F}_{des}*f . \label{eq:dfdt}\end{equation} We solve this
numerically using the conversion from $t$ to $z$ in a standard
$\Lambda$CDM Universe (see formulae in Elmegreen et al. 2004c).
The results are shown in Figure \ref{fig:deepbar2} for several
parameter values. The dotted line is the bar fraction as a
function of $z$ with only collisions operating and no destruction
using ${\cal I}_0=0.51$ Gy$^{-1}$. The dashed line is for internal
bar formation only, no destruction, and ${\cal F}_0=0.385$
Gy$^{-1}$. The dot-dashed line is for internal bar formation and
destruction with ${\cal F}_0=0.605$ Gy$^{-1}$ and ${\cal D}_0=1$.
All of these were tuned to give a bar fraction of 0.4 today, as
was the model for the top solid line, which has all three processes
acting together using ${\cal I}_0=0.51$ Gy$^{-1}$, ${\cal
F}_0=0.385$ Gy$^{-1}$, and ${\cal D}_0=1.85$.  Only the solid and
dotted lines are constant out to $z\sim1$. For the case with all
three processes, the time integral of the destruction rate, which
is the last term in equation \ref{eq:dfdt}, equals $-0.33$,
meaning that this fraction of all the galaxies had a bar
dissolve. The other solid line in figure \ref{fig:deepbar2} is
for a case with ${\cal I}_0=0.51$ Gy$^{-1}$, no internal formation
(${\cal F}_0=0$), and a constant destruction rate with ${\cal
F}_{des}=0.1$ Gy$^{-1}$.  This last example illustrates how even a
small amount of bar destruction (i.e., with a bar dissolution
timescale of 10 Gy) and no reformation gives a noticeably rising
bar fraction out to $z=1$.

\section{Conclusions}

Turbulence is related to galactic structure through the star
formation rate and morphology and through viscous forces and ISM
energy dissipation. ISM turbulence is a complicated process
involving thermal and magnetic pressures plus self-gravity, with
frequent and multi-scale energy input and dissipation. Between
bursts of energy input and when self-gravity is unimportant, MHD
turbulence has several properties that carry over from Kolmogorov
turbulence, including the power spectrum of the incompressible
part of the flow (the shear or solenoidal part).

Star formation is not just the result of gravitational collapse in
a turbulent medium because many sources of pressure, such as HII
regions and supernovae, make clouds independently of turbulence
and compress the turbulence-made clouds further, triggering
additional star formation. A high fraction of all star formation
may be triggered in this way. Inside cloud cores, turbulent
compression and self-gravity may dominate stellar compression.
Also, during the formation of spiral arms by gravitational
instabilities and the formation of giant molecular clouds by
turbulence and self-gravity, stellar pressures may be unimportant
because they are relatively rare on these scales.

Long-term disk evolution also depends on turbulence through its
effect on gas viscosity. Turbulent viscosity should be much
smaller than simulated viscosity with sticky particles unless the
particle collisions are frequent and highly dissipative. Turbulent
dissipation in the ISM is so rapid that the entire energy content
on the scale of the disk thickness has to be replaced every few
disk crossing times.  There are apparently enough energy sources
with close enough spacings to do this.

When strong bars or spirals are present, global disk accretion is
dominated by gravitational torques from these objects and by
torques at galactic shock fronts. Accretion by turbulent viscosity
is much slower. The appearance of bars at high $z$ with rather
normal abundance among disk galaxies suggests either that bars formed
early in the Universe and were not easily destroyed, as suggested by
Regan \& Teuben (2004), or that the bars which were destroyed were
promptly replaced by new bars, as suggested by Block et al.  (2002).
An important difference between these two simulations is the
treatment of viscosity.
The appearance of irregular galaxies with giant star-forming regions
suggests that turbulent velocities were large during the first Gigayear
in a galaxy's life.

\begin{acknowledgments}
B.G.E. acknowledges support from NSF Grant AST-0205097.
Helpful discussions with Debra Elmegreen are appreciated.
\end{acknowledgments}

\end{document}